\documentclass[11pt]{article}
\usepackage{jcapmod}

\usepackage{amsfonts, amsmath}
\usepackage{booktabs}
\usepackage[english]{babel}
\usepackage{colortbl}
\usepackage{graphicx}
\usepackage{subfig}
\usepackage{hyperref}
\usepackage{soul}
\usepackage{lipsum}
\usepackage{color}

\newcommand\blfootnote[1]{%
  \begingroup
  \renewcommand\thefootnote{}\footnote{#1}%
  \addtocounter{footnote}{-1}%
  \endgroup
}

\usepackage[load-configurations=astronomy, range-units=brackets, range-phrase=-, per-mode=reciprocal, mode=math]{siunitx}

\definecolor{pyBlue}{RGB}{31, 119, 180}
\definecolor{pyRed}{RGB}{214, 39, 40}
\definecolor{pyBlue2}{RGB}{0, 111, 237}
\definecolor{pyRed2}{RGB}{224, 52, 36}

\numberwithin{equation}{section}

\def\beq{\begin{equation}}
\def\eeq{\end{equation}}

\def\ell{l}
\def\fDM{{f_\mathrm{TCDM}}}

\DeclareSIUnit{\parsec}{pc}
\DeclareSIUnit{\Mpc}{\mega\parsec}
\DeclareSIUnit{\h}{\mathit{h}}

\DeclareRobustCommand{\SkipTocEntry}[4]{}

\begin{document}

\pagenumbering{roman}
\begin{titlepage}
	\baselineskip=15.5pt \thispagestyle{empty}
	
	\bigskip\
	
	\begin{center}
		{\fontsize{20.74}{24}\selectfont \sffamily \bfseries Dark Matter Interactions, \\ \vskip 10pt Helium, and the CMB}
	\end{center}
	
	\begin{center}
		{\fontsize{12}{30}\selectfont Roland de Putter,$^1$ Olivier Dor\'e,$^{1,2}$ J\'er\^ome Gleyzes,$^{1,2}$ Daniel Green,$^{3}$ Joel Meyers$^{4}$} 
	\end{center}
	
	\vspace{0.25cm}
	\begin{center}
		\textsl{$^1$California Institute of Technology, Pasadena, CA 91125}
        \vskip8pt
        \textsl{$^2$Jet Propulsion Laboratory, California Institute of Technology, Pasadena, CA, USA}
		\vskip8pt
		\textsl{$^3$ Department of Physics, University of California, San Diego, La Jolla, CA 92093, USA}
		\vskip8pt
        \textsl{$^4$ Canadian Institute for Theoretical Astrophysics, Toronto, ON M5S 3H8, Canada}

	\end{center}

	\vspace{1.2cm}
	\hrule \vspace{0.3cm}
	\noindent {\sffamily \bfseries Abstract}\\[0.1cm]
    The cosmic microwave background (CMB) places a variety of model-independent constraints on the strength interactions of the dominant component of dark matter with the Standard Model.  Percent-level subcomponents of the dark matter can evade the most stringent CMB bounds by mimicking the behavior of baryons, allowing for larger couplings and novel experimental signatures. However, in this note, we will show that such tightly coupled subcomponents leave a measurable imprint on the CMB that is well approximated by a change to the helium fraction, $Y_{\rm He}$.  Using the existing constraint on $Y_{\rm He}$, we derive a new upper limit on the fraction of tightly coupled dark matter, $f_{\rm TCDM}$, of $f_{\rm TCDM}<0.006$ (95\% C.I.).  We show that future CMB experiments can reach $f_{\rm TCDM}<0.001$ (95\%  C.I.) and confirm that the bounds derived in this way agree with the results of a complete analysis.  We briefly comment on the implications for model building, including milli-charged dark matter.\blfootnote{\textcircled{c} 2018. All rights reserved. } 
    
	\vskip10pt
	\hrule
	\vskip10pt

\end{titlepage}


\clearpage
\pagenumbering{arabic}
\setcounter{page}{1}

\clearpage
\section{Introduction}
\label{sec:introduction}

The nature of dark matter is one of the central questions in physics, intersecting the fields of astrophysics, cosmology, and particle physics.  While it is common (and simpler) to assume dark matter is made entirely from a single type of weakly interacting particles, the possibility that the dark sector involves a richer set of particles and interactions is compatible with (or even motivated by) the state of experimental and observational dark matter searches (see e.g.,~\cite{Lisanti:2016jxe,Slatyer:2017sev} for recent reviews).  A variety of new searches have been proposed that target this type of exotic physics in the dark sector~\cite{Battaglieri:2017aum}.

Cosmology provides a particularly useful window into the nature of dark matter.  It is the domain where the influence of dark matter can be inferred most directly and unambiguously from observations.  In the linear regime, predictions for the evolution of the density fluctuations in the presence of cold dark matter (CDM) have been verified at high precision with cosmological observations~\cite{Ade:2015xua}.  These observations put strong limits on potential interactions between the dominant component of dark matter and the Standard Model~\cite{Dvorkin:2013cea,Gluscevic:2017ywp,Boddy:2018kfv} and/or other particles in the dark sector (e.g.,~\cite{Buen-Abad:2017gxg,Pan:2018zha,Raveri:2017jto}).

The cosmological bounds on these interactions are drastically reduced for percent-level subcomponents of the dark matter as they can mimic the signatures of baryons~\cite{Dubovsky:2003yn,Dolgov:2013una}. Baryons make up roughly five percent of the energy density in the universe today and have substantial self-interactions and interactions with the photons.  Since it is these interactions that distinguish baryons from the dark matter in many cosmological observables, a subcomponent of dark matter strongly interacting with the Standard Model is effectively a baryon from the cosmological perspective.  These limitations on cosmological constraining power are particularly relevant to light dark matter subcomponents that interact with baryons either through electromagnetism (e.g. milli-charge) or some new direct force.  These examples are not well constrained by current experiments but can have observable signatures in a variety of proposed searches~\cite{Battaglieri:2017aum}.  These models have also been proposed to explain unusual astrophysical signals like the EDGES measurement of the global 21cm absorption feature~\cite{Bowman:2018yin,Barkana:2018lgd,Munoz:2018pzp}.

Despite the natural intrinsic degeneracy between baryons and tightly coupled dark matter (TCDM), we will show in this note that cosmic microwave background (CMB) signatures of TCDM subcomponents are distinguishable from a change to the abundance of baryons.
At the time of recombination, the phenomenology of TCDM is effectively the same as a change to the primordial helium abundance, $Y_{\rm He}$, as far as the CMB is concerned: neutral helium does not efficiently interact with CMB photons, yet it is tightly coupled to protons and electrons and contributes to the physical baryon density, $\omega_b$. At fixed total effective baryon density, increasing the mass fraction in the form of helium or TCDM therefore increases the mean free path of photons by decreasing the density of scatterers which, in turn, increases diffusion (Silk) damping of the power spectrum\footnote{A suppression of small scale power was noted numerically in~\cite{Dubovsky:2003yn,Dolgov:2013una} in the context of milli-charged dark matter.  Here we explain the physical origin of this suppression and its analytic form, and use this understanding to derive a new constraint on the fraction of tightly coupled dark matter.}. From this observation, the CMB limits on the fraction of TCDM can be derived directly from the existing CMB measurement of $Y_{\rm He}$ and does not require a separate analysis of the data.  Furthermore, this bound does not depend on the detailed form of the interactions between dark matter and the baryons\footnote{We are using the cosmological convention where the electrons are included as a component of the baryons.}. With current limits on $Y_{\rm He}$ from Planck~\cite{Ade:2015xua}, we will show that TCDM can be at most 0.6\% of the dark matter, and upcoming CMB observations should improve these limits by a factor of five.  This new bound excludes the most of the viable parameter space for a milli-charged dark matter interpretation of the EDGES measurement~\cite{Berlin:2018sjs,Munoz:2018pzp} and is similarly powerful in other settings as a broad model-independent constraint.  

\section{Tight Coupling and Helium}

In this note, we will assume that the dark matter has sufficiently large interactions with the baryons to be tightly coupled with the photon-baryon fluid during the era of recombination. We will also assume its direct scattering cross-section with the photons to be negligible as to not alter the visibility function.  These assumptions are relevant to direct forces between baryons and the dark matter, for example due to the exchange of a new particle or a milli-charge\footnote{For a wide range of milli-charges, milli-charged dark matter is tightly coupled to the baryons, while the scattering between milli-charged dark matter and photons is negligible. The reason is that Compton scattering of photons is suppressed by two additional powers of the milli-charge relative to the Coulomb interaction between the dark matter and the baryons.
}~\cite{Holdom:1985ag}.

Our definition of tight coupling is that the TCDM component and baryons behave effectively as a single fluid.  Assuming the dark matter is non-relativistic during recombination (mass greater than 10 eV), its contribution to this fluid is only to increase the energy density. Prior to recombination, the baryon-TCDM fluid couples to the photons to form a single relativistic fluid with a sound speed,
\beq
c_s = \frac{1}{\sqrt{3(1+R_b)}}  \qquad R_b \equiv \frac{3 \rho_b}{ 4 \rho_\gamma} \ ,
\eeq
leading to characteristic temperature fluctuations, $\Delta = \Delta T/T$, of the schematic form,
\beq
(\Delta +\psi)(k, \eta) \propto \cos(k \, r_s(\eta)) \qquad r_s(\eta) = \int^\eta d\eta' c_s(\eta') \ ,
\eeq
where $\psi$ is the Newtonian gauge (time-time) metric perturbation.
Since the pressure comes only from the photons, increasing the effective baryon density (either in the form of true baryons or TCDM) at fixed photon density will decrease the sound speed.  After recombination, the neutral hydrogen is decoupled from the photons and evolves similarly to the TCDM up until reioniziation.

From this description, one might imagine that there is no discernible difference between TCDM and true baryon density as measured by the CMB power spectrum.  Indeed the first cosmological bounds on the fraction of TCDM~\cite{Dubovsky:2003yn} came from comparing the measurement of $\omega_b$ inferred from the CMB (with $Y_{\rm He}$ fixed) to that from primordial light element abundances produced during big bang nucleosynthesis (BBN). TCDM does not contribute to $\omega_b$ measured from BBN and therefore the difference is a measure of the TCDM fraction.

Fortunately, we can get a better constraint from the CMB alone. While baryons and TCDM are indistinguishable both well before recombination when baryons and photons are tightly coupled as well as after decoupling (but before reionization), the degeneracy is broken (or modified) by the finite mean free path of photons in the intermediate period, before full decoupling.  Specifically, photons and baryons do not form a genuinely perfect fluid due to photon diffusion out of the gravitational potential wells formed predominantly by the dark matter.  This diffusion dissipates the energy of the sound waves in the fluid and leads to the exponential damping of a given mode, 
\beq
(\Delta +\psi)(\vec k, \eta) =A_{\vec k} \cos(k \, r_s(\eta)) e^{- k^2 / k_d^2} \ ,
\eeq
where $A_{\vec k}$ is the primordial amplitude.  The diffusion (Silk) damping scale, $k_d$, is given by~\cite{Silk:1967aha,Weinberg:1971mx,Kaiser:1983abc}
\beq
\frac{1}{k_d^2(\eta)} = \int^\eta d\eta' \frac{1}{a n_e x_e \sigma_T}  \frac{c_s^2}{2} \left[ \frac{R_b^2}{(1+R_b)} + \frac{16}{15} \right] \label{eq:k_d}
\eeq
where $\sigma_T$ is the Thomson scattering cross-section, $n_e$ is the number-density of electrons, and $x_e$ is the ionization fraction\footnote{We will use a common convention where $n_e$ ignores electrons bound in helium such that neutrality requires $n_e =n_H$ where $n_H$ is the number density of free protons.  This definition implies that $x_e$ can exceed unity when helium is ionized.}. 

One noteworthy feature of the damping scale $k_d$ (and the mean free path in general) is that it is determined by $n_e$, which depends both on the effective baryon density and the form of the baryons. Charge neutrality of the universe requires that $n_e = n_{H}$ where $n_H \neq n_b$ is the number density of hydrogen nuclei (free protons).  This distinction is important even in $\Lambda$CDM due to the presence of primordial helium. At the time of recombination, helium is tightly coupled to the protons but is electrically neutral (not ionized) and does not efficiently couple to CMB photons.  As a result, at fixed $\omega_b\equiv \Omega_b h^2$, the damping scale is sensitive to the helium mass fraction, $Y_{\rm He}$, through $n_e = n_b (1- Y_{\rm He})$.  It is predominantly through this effect that the CMB can constrain $Y_{\rm He}$, leading to the current measurement \cite{Ade:2015xua},
\beq
\label{eq:YHe meas}
Y_{\rm He} = 0.249^{+0.026}_{-0.027} \quad (95\% \,\text{C.I.}),
\eeq
assuming a $\Lambda$CDM + $Y_{\rm He}$ cosmology.

Coming back to TCDM, the above discussion shows that, instead of $\omega_{\rm TCDM}$ being degenerate with $\omega_b$ (with $Y_{\rm He}$ fixed by BBN consistency), $\omega_{\rm TCDM}$ is really degenerate with the helium density. We explain below how we can therefore use the CMB constraint on $Y_{\rm He}$ of Eq.~(\ref{eq:YHe meas}), in combination with the $Y_{\rm He}$ value predicted by BBN, to derive constraint on the TCDM fraction.

In the presence of TCDM, we will simply define 
\beq
n_b \equiv \frac{\rho_b}{m_{\rm H}} \equiv n_{\rm H} +\frac{m_{\rm He}}{m_{\rm H}} n_{\rm He} + \frac{\rho_{\rm TCDM}}{ m_{\rm H}} \ ,
\eeq
such that $\rho_b$ and $n_b$ include the TCDM.  Even in the presence of TCDM, we will still define $Y_{\rm He}$ in terms of the true baryons as they are the relevant quantity during BBN,
\beq
Y_{\rm He} \equiv \frac{m_{\rm He} n_{\rm He}}{m_{\rm H}n_{\rm H} + m_{\rm He} n_{\rm He}} = \left(1-\frac{\omega_{\rm TCDM}}{\omega_{b}} \right)^{-1}\frac{m_{\rm He}}{m_{\rm H}} \, \frac{ n_{\rm He}}{  n_b} \ .
\eeq
At fixed $\omega_b$, the number density of electrons is therefore
\beq
n_e = n_b \left[ 1 - Y_{\rm He} - (1-Y_{\rm He})\frac{\omega_{\rm TCDM}}{\omega_{b}} \right] \equiv n_b \left[ 1 - Y_{\rm He} -F_{\rm TCDM} \right]\ . \label{eq:n_e} 
\eeq
This formula defines $F_{\rm TCDM}$ as it is only through this parameter that $\omega_{\rm TCDM}$ enters the Boltzmann equations.

It should be clear that a change to $Y_{\rm He}$ or $F_{\rm TCDM}$ (while holding the other fixed) has the same effect on Eq.~(\ref{eq:n_e}). Using the BBN prediction of $Y_{\rm He} = 0.24534 \pm 0.00061$ and measured $\omega_b /\omega_{\rm cdm} = 0.185 \pm 0.004$~\cite{Ade:2015xua}, we can therefore translate the measurement of $Y_{\rm He}$ in Eq.~(\ref{eq:YHe meas}) to a constraint
\beq
f_{\rm TCDM} \equiv \frac{\omega_{\rm TCDM}}{\omega_{\rm cdm}} < 0.0070 \quad (95\% \, {\rm C.I.}) 
\label{eq:fTCDM_constraint} \ .
\eeq
This upper limit of approximately 0.7\% tightly coupled dark matter is almost 10 percent weaker than the true bound because it includes a change to $x_e(z)$ in Eq.~(\ref{eq:k_d}) due to helium ionization that is not associated with TCDM. We will demonstrate this small additional effect in the next section and confirm that extrapolation of the limits on $Y_{\rm He}$ is consistent with the treatment of TCDM in a full Boltzmann code.  In addition, we will see that significant improvements in these bounds are expected in the next generation of experiments.  The bound derived here is already an improvement over the best previously published constraint~\cite{Dolgov:2013una} (where the TCDM was assumed to be milli-charged dark matter) from the inclusion of Planck polarization data\footnote{Using Planck TT-only measurement, $Y_{\rm He} = 0.251^{+0.041}_{-0.042}$ (95\% C.I.), one finds $f_{\rm TCDM} < 0.011$ (95\% C.I.) which is consistent with the previous upper limit on the fraction of milli-charged dark matter, $\omega_{\rm MCDM} < 0.001$ (95\% C.I.), found in~\cite{Dolgov:2013una}.}.
 
This bound is particularly relevant to dark matter-baryon interactions invoked to explain the EDGES measurement~\cite{Bowman:2018yin,Barkana:2018lgd}.  Given existing experimental and observational constraints, the most viable such model is a tightly coupled milli-charged subcomponent of the dark matter with $f_{\rm TCDM} \sim 0.003-0.01$~\cite{Berlin:2018sjs,Munoz:2018pzp}.  Our new bound significantly reduces the available parameter space for these models and should be covered entirely by the current generation of ground-based CMB observations.

The bounds presented here also have implications to experimental searches for light dark matter.  While direct detection searches have placed strong limits on the dark matter-baryon cross sections for masses above a GeV, few experiments directly probe smaller masses (see e.g.~\cite{Essig:2013lka,Battaglieri:2017aum} for reviews).  A  variety of indirect measurements limit the allowed parameter space for the dominant component of dark matter~\cite{Ade:2015xua,Green:2017ybv,Knapen:2017xzo,Boddy:2018kfv}. These indirect constraints can be weakened significantly for subcomponents of the dark matter and leave open the opportunity of direct detection in the lab.  The limits presented here further restrict the viable parameter space for such models.

\begin{figure}
\begin{centering}
\includegraphics[width=0.9\columnwidth]{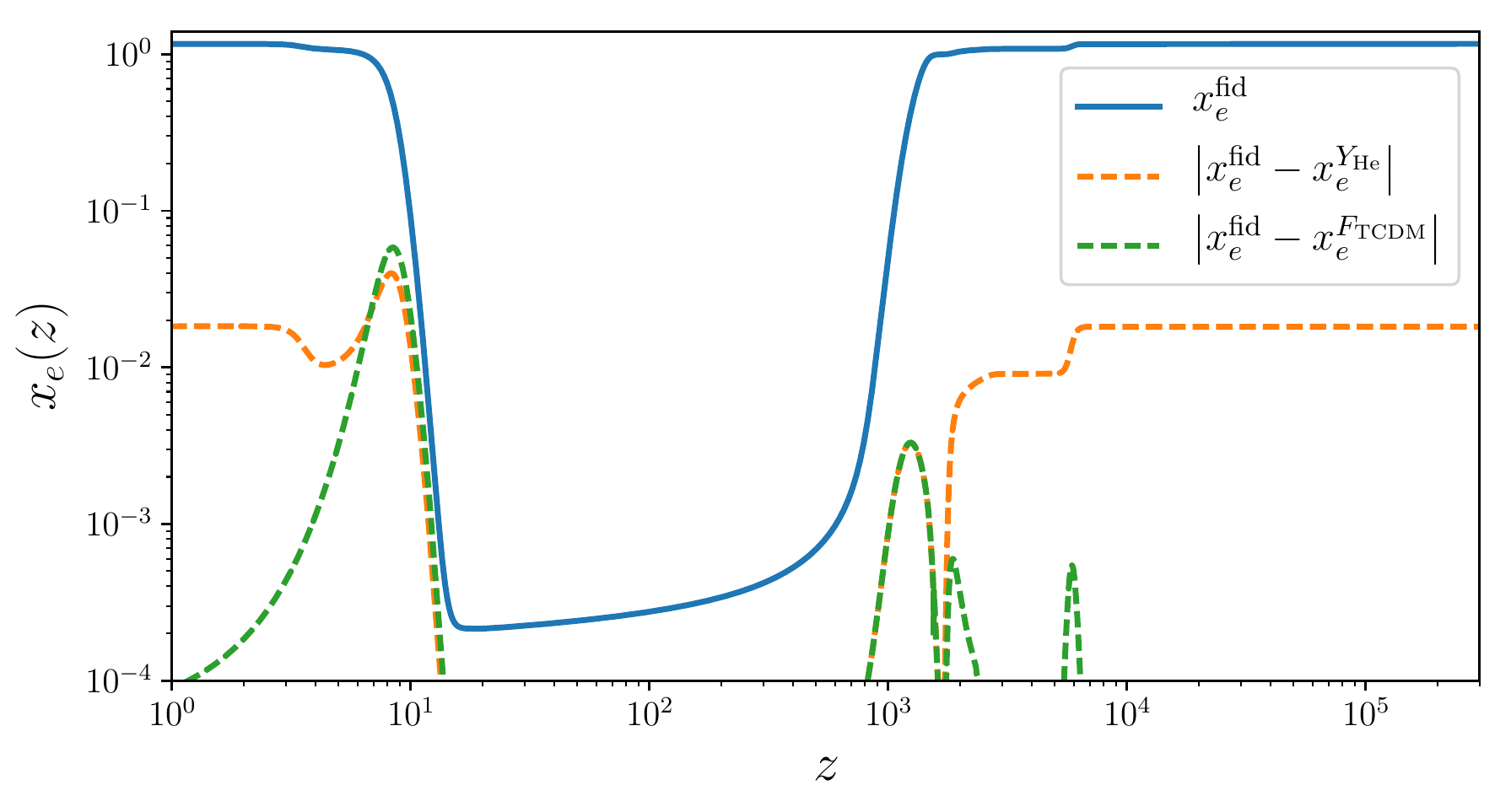}
\caption{ Changes to the ionization fraction resulting from an increase of 0.02 to $Y_{\rm He}$ and to $F_{\rm TCDM}$.  The dominant effects around hydrogen recombination are almost identical for both parameters. First and second helium recombination are visible in the orange curve as rapid changes in $x_e(z)$ around $z \simeq$ 6000 and 1800 and likewise for helium reionization around $z\simeq 3$.  These features are easily visible when we change $Y_{\rm He}$ but are not affected by changes to $F_{\rm TCDM}$. }
\label{fig:xe_history}
\end{centering}
\end{figure}

\begin{figure}
\begin{centering}
\includegraphics[width=0.9\columnwidth]{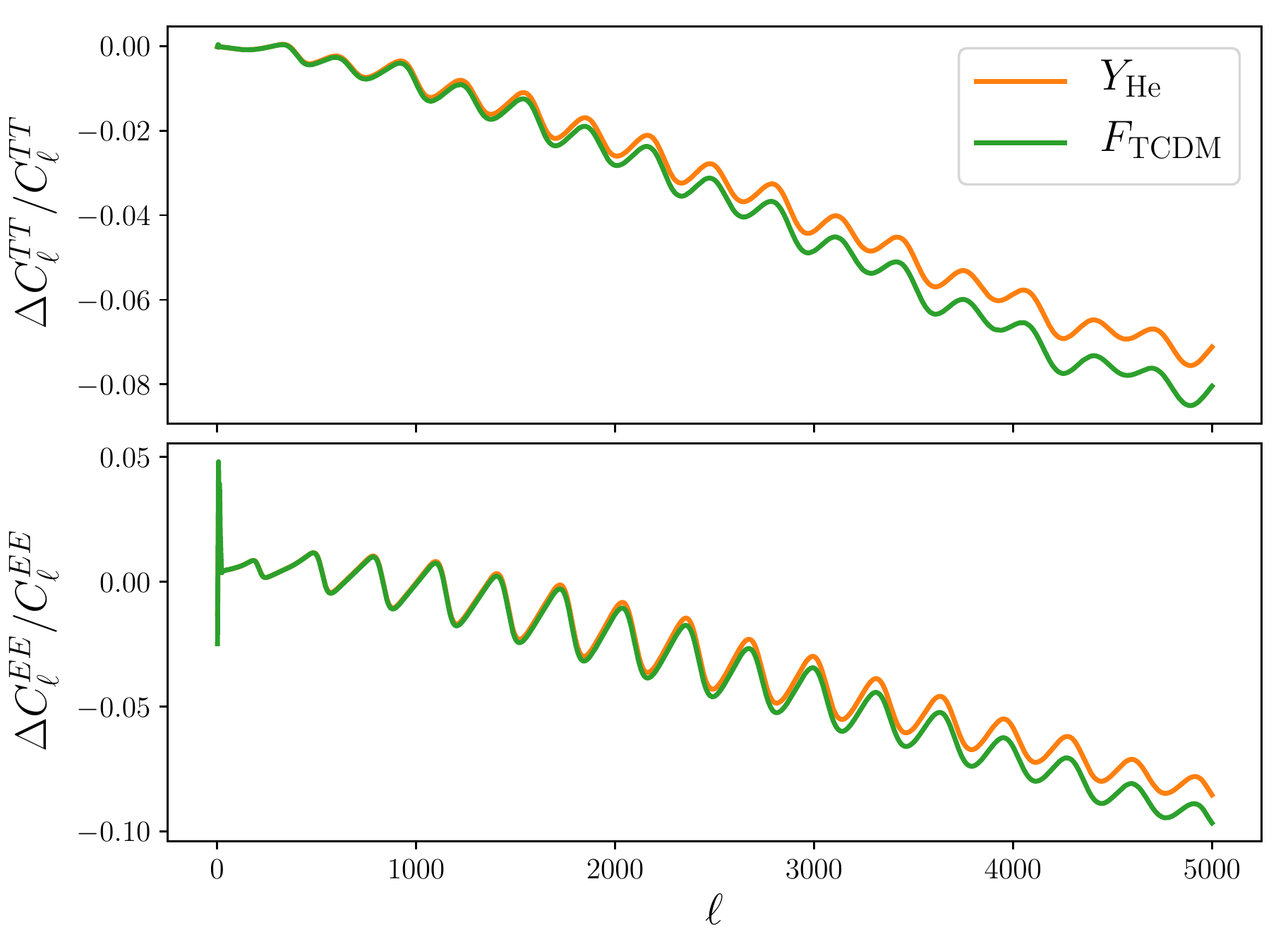}
\caption{ Relative changes to the temperature (top) and E-mode polarization (bottom) power spectra resulting from an increase of 0.02 to $Y_{\rm He}$ and $F_{\rm TCDM}$.}
\label{fig:TTEE}
\end{centering}
\end{figure}

\begin{figure}
\begin{centering}
\includegraphics[width=0.9\columnwidth]{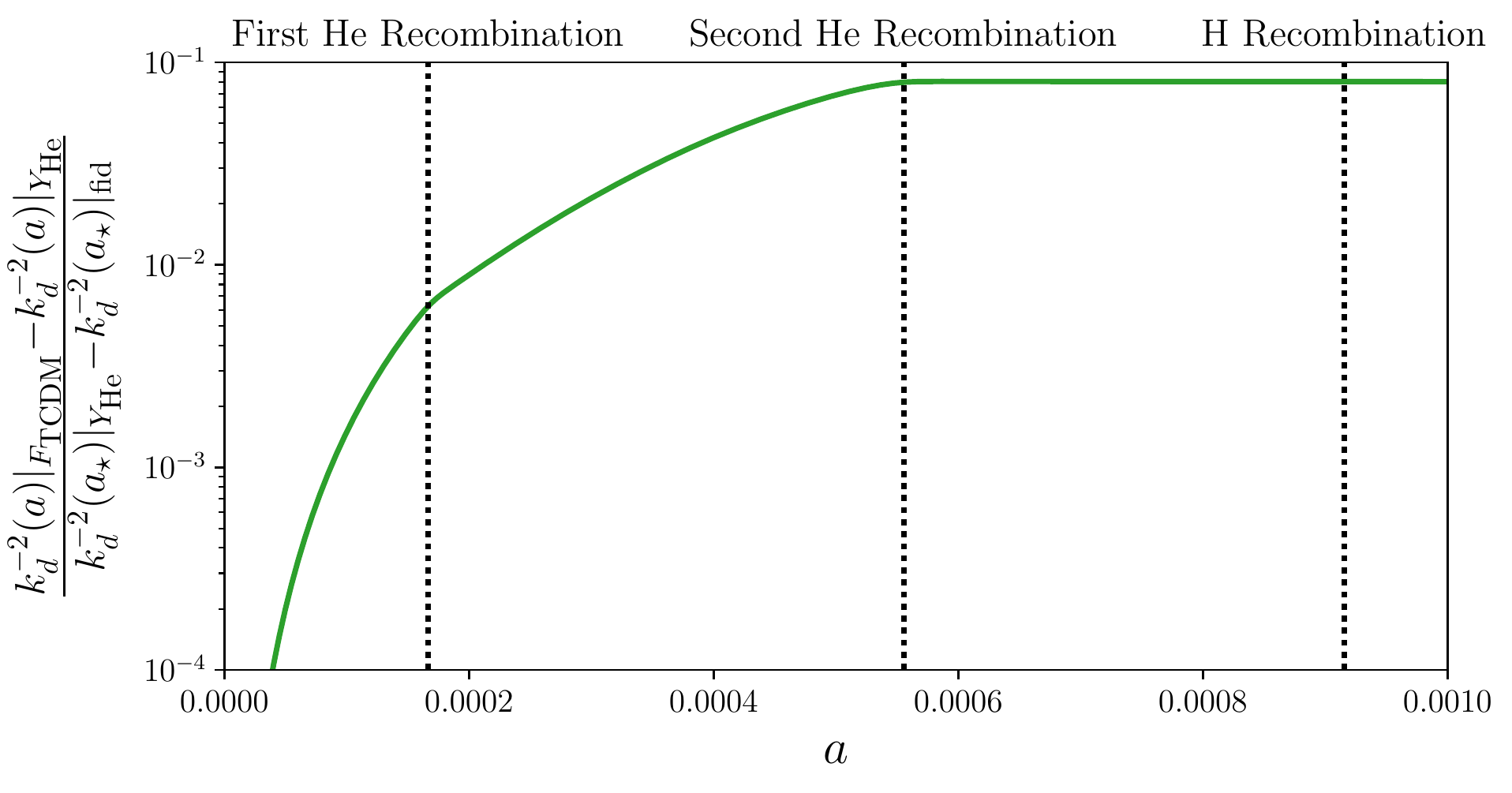}
\caption{ Difference in $k_d^{-2}(a)$ (see Eq.~\eqref{eq:k_d}) between models where $Y_{\rm He}$ and $F_{\rm TCDM}$ are increased by 0.02 relative to the difference in $k_d^{-2}(a_\star)$ between a model with $Y_{\rm He}$ increased by 0.02 and the fiducial model, where $a_\star$ is the scale factor at last scattering.  It is clear that the difference in the damping scale accumulates during the very early times when helium has not yet recombined and the value of the ionization fraction $x_e(z)$ differs between the two models.  The difference reaches a constant 8\% offset after helium recombination. }
\label{fig:rD2}
\end{centering}
\end{figure}

\begin{figure}
\begin{centering}
\includegraphics[width=0.75 \columnwidth]{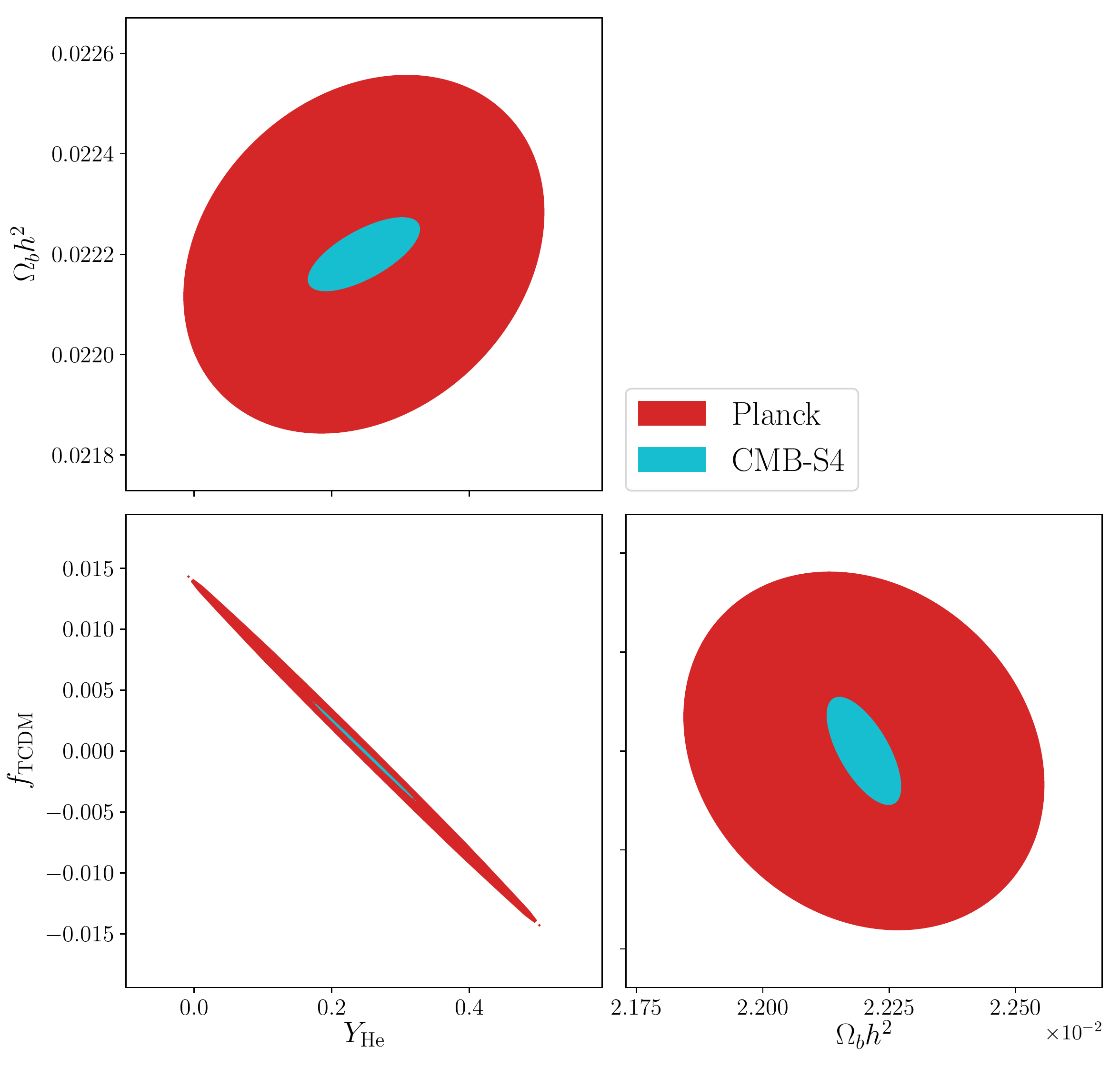}
\caption{ Forecasts for joint 1$\sigma$ constraints on $Y_{\rm He}$ and $\fDM$ showing the significant degeneracy that exists between the two parameters.  We show forecasts for noise levels similar to current Planck data and the forthcoming CMB-S4 survey~\cite{Abazajian:2016yjj}.  We include unphysical negative values of $Y_{\rm He}$ and $\fDM$ to clearly display the striking degeneracy. The direction of the degeneracy is consistent with the scaling shown in Eq.~\eqref{eq:scaling}.}
\label{fig:fisher_ellipse}
\end{centering}
\end{figure}

\begin{figure}
\begin{centering}
\includegraphics[width=0.9\columnwidth]{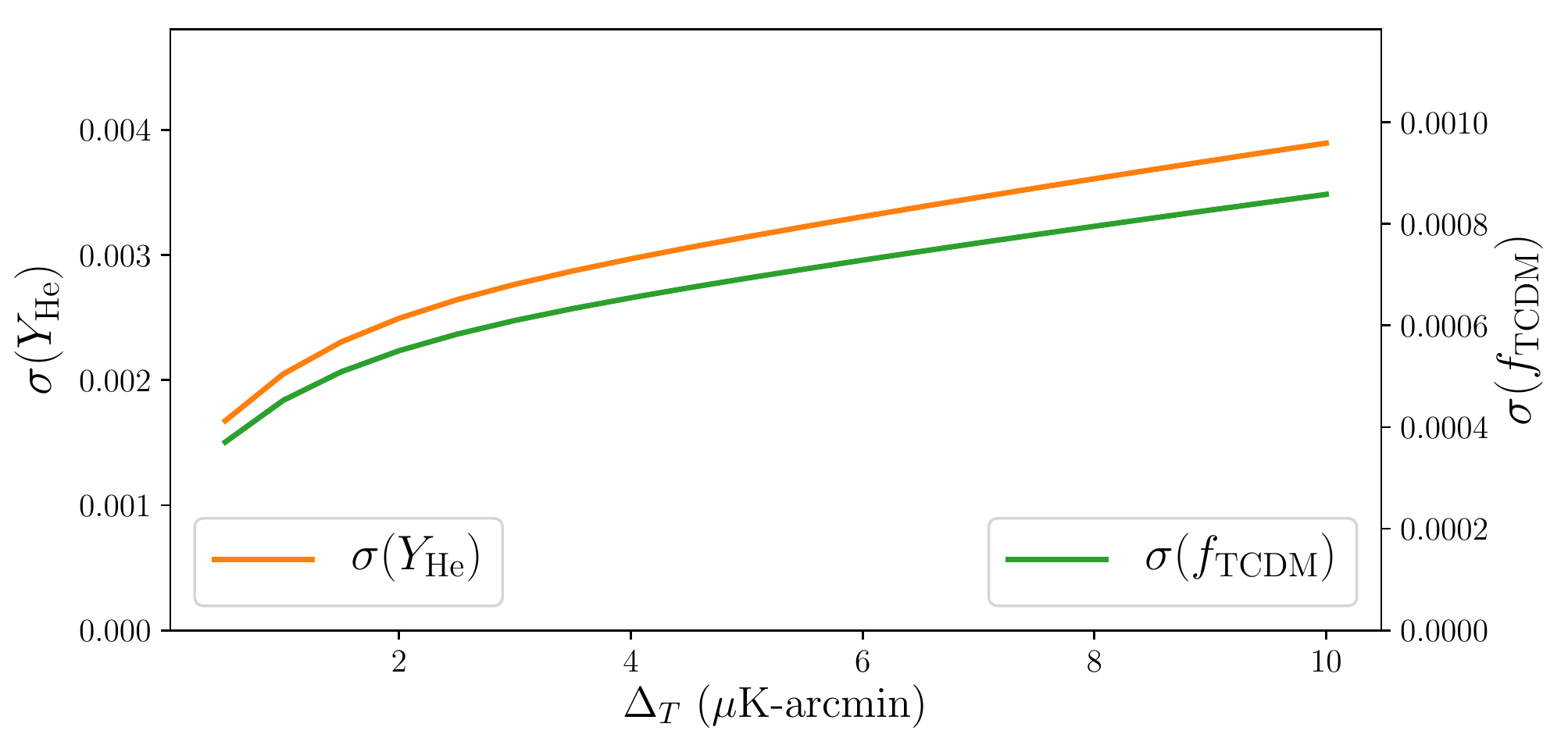}
\caption{ Forecasts for $1\sigma$ constraints on $Y_{\rm He}$ in $\Lambda$CDM+$Y_{\rm He}$ compared to constraints on $\fDM$ in $\Lambda$CDM+$\fDM$ for a range of noise levels, assuming a 1.5 arcminute beam.  Our forecasts show that to a very good approximation the direct constraints on tightly coupled dark matter can be obtained from rescaling constraints on the primordial helium abundance.  These two forecasts differ by a nearly constant factor $\sigma(f_{\rm TCDM}) \approx 0.90 \  (\omega_b/ \omega_{\rm cdm}) \ (1-Y_{\rm He})^{-1} \ \sigma(Y_{\rm He})$, where the additional factor of 0.90 is due to the increase in $x_e$ at high redshifts from helium ionization.}
\label{fig:forecasts}
\end{centering}
\end{figure}

\section{Validation and Forecasts}

The above physical argument suggests that the identification between $F_{\rm TCDM}$ and $Y_{\rm He}$ should be valid for a CMB power spectrum analysis.  However, the constraints on $Y_{\rm He}$ are derived using a Boltzmann code like \texttt{CAMB}~\cite{Lewis:1999bs} or \texttt{CLASS}~\cite{Blas:2011rf} to compute the CMB power spectrum, and it is reasonable to ask if this identification is valid at the level of precision needed both in current and future data.

The identification between $F_{\rm TCDM}$ and $Y_{\rm He}$ would seem to be exact with respect to the Boltzmann equations for the evolution of the photon and baryon density perturbations.  Helium is included only as a subcomponent of the baryon density and is not given its own evolution equations.  For fixed $\omega_b$ and ionization history $x_e(z)$, the only impact of $Y_{\rm He}$ is to change $n_e$ as described above.
In this sense, the evolution of the perturbations treats helium and TCDM in the same way.

However, there is more to the cosmology of helium than simply altering $n_e$ at recombination.  In particular, the ionization history\footnote{The evolution of the gas temperature is also sensitive to helium but has negligible observable impact on the CMB. }, as defined by $x_e(z)$, must include helium ionization both in the early and late universe.  The atomic physics of helium is included when computing the ionization/recombination history~\cite{Seager:1999bc,AliHaimoud:2010dx,Chluba:2010ca}. From the output of such a calculation, the distinction between helium and TCDM can be easily seen from the fact that one finds redshifts where $x_e(z)>1$ in the presence of (ionized) helium.  As shown in Figure~\ref{fig:xe_history}, the response of $x_e(z)$ to changes in $Y_{\rm He}$ and $F_{\rm TCDM}$ is nearly identical around recombination where $x_e$ is responding only to the change in $n_e$.  In contrast, the results are significantly different at high and low redshifts where helium ionization is important.

CMB observations do not directly constrain $x_e(z)$ and we must therefore address whether this difference between TCDM and helium impacts the constraints on $\fDM$ derived from constraints on $Y_{\rm He}$. In order to validate the assumption made in the previous section, we modified \texttt{CAMB}~\cite{Lewis:1999bs} to include a TCDM component.  This was achieved by reducing the number density of protons (and thereby electrons) compared to the density of effective baryons as shown in Eq.~\eqref{eq:n_e}.  The density of baryons used in the recombination and reionization parts of the code was adjusted to match the true baryon density (rather than the effective density which includes the TCDM).  With these modifications an equal change to either $Y_{\rm He}$ or $F_{\rm TCDM}$ leads to an ionization history which is the same when helium is neutral, but differs before helium recombination and after helium reionization.

The effects on the CMB power spectra for equal changes to $Y_{\rm He}$ and $F_{\rm TCDM}$ are very similar, as seen in Figure~\ref{fig:TTEE}.  There is a slight difference in the amount of damping, which can be straightforwardly understood as being due to the presence of ionized helium at high redshift. Models with increased $Y_{\rm He}$ have higher $x_e(z)$ at very early times before helium recombined.   Helium ionization adds additional electrons, which reduces the mean free path of photons at higher redshifts and thus increases $k_d$, as can be seen in Eq.~\eqref{eq:k_d} and in Figure~\ref{fig:rD2}. Directly calculating $k_d$ gives an 8\% difference in the change to $k_d^2$ between $Y_{\rm He}$ and $F_{\rm TCDM}$ relative to the fiducial model.  

There is also a difference in the way reionization proceeds when either $Y_{\rm He}$ or $F_{\rm TCDM}$ is adjusted.  Our modifications held fixed the optical depth to reionization, $\tau$, which requires a slightly different redshift of reionization in the two cases, since helium undergoes reionization while TCDM does not.  This difference is reflected in the impact on the largest scales of the $E$-mode polarization power spectrum, shown in the bottom panel of   Figure~\ref{fig:TTEE}, though these modes are subject to large cosmic variance and do not drive the constraints shown below.  The response of $x_e(z)$ to changes in $Y_{\rm He}$ and $F_{\rm TCDM}$ also alters the visibility function during recombination but we find this effect is much smaller than the change to $k_d$.

The similarities seen in the ionization history and the power spectra translate directly to the measurements of $Y_{\rm He}$ and $\fDM$.  Forecasts including both parameters show a remarkable degeneracy, as shown in Figure~\ref{fig:fisher_ellipse}.  This shows that their impact on CMB power spectra is essentially identical.  To provide a more quantitative comparison, we show the forecasts for $\Lambda$CDM+$Y_{\rm He}$ and $\Lambda$CDM+$\fDM$ in Figure~\ref{fig:forecasts}.  Given our analytic argument, rescaling $Y_{\rm He}$ forecasts by a factor of $(\omega_b/\omega_{\rm cdm}) \ (1-Y_{\rm He})^{-1}$ should produce the nearly same results.  The forecasts show that the two agree very well, but in detail
\beq 
\sigma(f_{\rm TCDM}) \approx 0.90 \  (\omega_b/ \omega_{\rm cdm}) \ (1-Y_{\rm He})^{-1} \ \sigma(Y_{\rm He}) \, .
\label{eq:scaling}
\eeq 
This additional factor of 0.90 is related to the 8\% difference in $k_d$ arising from the helium ionization at early times through the factor of $x_e(z)$ in Eq.~\eqref{eq:k_d}. Since $F_{\rm TCDM}$ leads to slightly more damping than $Y_{\rm He}$, the constraint on $\fDM$ derived from $Y_{\rm He}$ is weaker than the true bound by 10\%.   For completeness, we computed the actual bound using the Planck 2015 likelihood~\cite{Aghanim:2015xee} and a modified version of \texttt{CosmoMC}~\cite{Lewis:2002ah} to find \beq
\fDM < 0.0064 \quad (95\% {\rm C.I.}) \, ,
\eeq
which very nearly matches 0.90 times the constraint quoted in Eq.~\eqref{eq:fTCDM_constraint} obtained from rescaling the $Y_{\rm He}$ constraint.  Since the ratio of the constraints is largely independent of noise level, one could further rescale the bounds of the previous section to give limits accurate to about a percent.

Finally, the forecasts shown in  Figure~\ref{fig:forecasts} illustrate the significant improvement in sensitivity to TCDM expected for the coming generations of CMB surveys.  A survey like CMB-S4~\cite{Abazajian:2016yjj} with map noise levels near 1$\mu$K-arcmin can reach $f_{\rm TCDM}<0.001$ (95\%  C.I.).  This factor of 5-6 improvement is unsurprising given the relationship to $Y_{\rm He}$ we have shown here, but it further illustrates that the degeneracy with $\omega_b$ does not fundamentally limit our sensitivity to TCDM.

\section{Summary}

Subcomponents of dark matter which interact with baryons can evade some of the stringent model-independent constraints placed by cosmology.  These subcomponents are appealing as experimental targets and signatures of complex dark sectors.  In this note, we showed that tightly coupled subcomponents of dark matter would effectively increase the helium fraction as measured by the CMB and can be excluded by current Planck data if they make up more than 0.6\% of the dark matter.  Upcoming CMB observations will improve on this bound by about a factor of five.

These results are relevant to the broader question of how cosmological data more generally informs our understanding of dark matter.  It has long been understood that the CMB measurement of $Y_{\rm He}$ will continue to improve with observational sensitivity~\cite{Abazajian:2016yjj}.  However, the potential impact of these measurements was largely thought to be restricted to probing BBN, which is already well characterized by primordial abundance measurements.  Nonetheless, we have shown that, unlike primordial abundances, the CMB constraint is not particularly sensitive to the atomic or nuclear properties of helium and can broadly characterize new physics which does not directly impact BBN.  This further suggests that combining primordial abundance measurements and CMB observations may yield further insights into the nature of dark matter.

More generally, the tools we use to characterize dark matter from cosmology are being significantly extended with improvements in CMB sensitivity.  Although the signatures of tightly coupled dark matter are degenerate with $\omega_b$ to leading order, the sub-leading impact on the damping tail offers a powerful constraint.  As intuition for cosmological data evolves to match the exponential improvements in survey sensitivity, new windows into the nature of dark matter are likely to emerge.

\vskip20pt
\paragraph{Acknowledgements}
We thank Yacine Ali-Ha\"imoud, Kimberly Boddy, and Surjeet Rajendran for helpful conversations.  We acknowledge the use of \texttt{CAMB}~\cite{Lewis:1999bs}, \texttt{CLASS}~\cite{Blas:2011rf}, and \texttt{CosmoMC}~\cite{Lewis:2002ah}. Part of the research described in this paper was carried out at the Jet Propulsion Laboratory, California Institute of Technology, under a contract with the National Aeronautics and Space Administration. R.d.P. and O.D. acknowledge the generous support from the Heising-Simons Foundation.

\clearpage
\phantomsection
\addcontentsline{toc}{section}{References}
\bibliographystyle{utphys}
\bibliography{dark_matter_Yp}

\end{document}